\documentstyle[12pt]{article}

\newcommand{\eq}{\begin{equation}}
\newcommand{\en}{\end{equation}}

\newcommand{\eqa}{\begin{eqnarray}}
\newcommand{\ena}{\end{eqnarray}}
\newcommand{\eqan}{\begin{eqnarray*}}
\newcommand{\enan}{\end{eqnarray*}}

\newcommand{\lbl}{\label}

\newcommand{\JAP}[1]{J. Appl. Phys.\ {\bf #1}\ }
\newcommand{\JMP}[1]{J. Mod. Phys.\ {\bf #1}\ }
\newcommand{\JP}[1]{J. Phys.\ {\bf #1}\ }

\newcommand{\PL}[1]{Phys. Lett.\ {\bf #1}\ }
\newcommand{\PR}[1]{Phys. Rev\ {\bf #1}\ }

\newcommand{\RNC}[1]{Riv. Nuovo Cimento\ {\bf #1}\ }

\def\sqr#1#2{{\vcenter{\hrule height.#2pt
     \hbox{\vrule width.#2pt height#1pt \kern#1pt
        \vrule width.#2pt}
     \hrule height.#2pt}}}

\def\thinspace{\kern .16667em}
\def\Dir{\nabla\kern-7.8pt\Big{/}}

\def\reali{{\hbox{\s@ l\kern-.5mm R}}}
\def\naturali{{\hbox{\s@ l\kern-.5mm N}}}
\def\interi{{\mathchoice
 {\hbox{Z\kern-1.5mm Z}}
 {\hbox{Z\kern-1.5mm Z}}
 {\hbox{{Z\kern-1.2mm Z}}}
 {\hbox{{Z\kern-1.2mm Z}}}  }}

\def\unity{{\hbox{\s@ 1\kern-.8mm l}}}
\def\uno{{\hbox{ 1\kern-.8mm l}}}
\def\part{\partial}

\def\aa{\alpha}
\def\bb{\beta}

\def\dd{\delta}

\def\ee{\epsilon}
\def\eb{\bar\epsilon}

\def\ff{\phi}
\def\fs{\phi^{\star}}
\def\fd{\phi^{\dagger}}

\def\vf{\varphi}

\def\GG{\Gamma}

\def\tt{\theta}

\def\G{{\cal G}}

\begin{document}
\begin{titlepage}
\begin{flushright}
NBI-HE-93-07\\
January 1993\\
PRELIMINARY\\
hep-th/9302007
\end{flushright}
\vspace*{0.5cm}
\begin{center}
{\bf
\begin{Large}
{\bf
VECTOR INDUCED LATTICE GAUGE THEORIES\\}
\end{Large}
}
\vspace*{1.5cm}
         {\large Igor Pesando}\footnote{E-mail PESANDO@NBIVAX.NBI.DK,
22105::PESANDO, 39163::I\_PESANDO}
         \\[.5cm]
%         Dipartimento di Fisica Teorica dell'Universit\`{a} di Torino
          The Niels Bohr Institute\\
          University of Copenhagen\\
          Blegdamsvej 17, DK-2100 Copenhagen \O \\
          Denmark
\end{center}
\vspace*{0.7cm}
\begin{abstract}
{
We consider vector induced lattice gauge theories,
in particular we consider the QED induced and we show that
at negative temperature corresponds to the dimer problem
while at positive temperature it describes a gas of branched
polymers with loops.
We argue that these models are critical at $D_{cr}=6$ for $N\ne\infty$
and they are {\sl not} critical for $N=\infty$.
}
\end{abstract}
\vfill
\end{titlepage}

\setcounter{footnote}{0}
\def\ut{{\tilde u}}
\def\zt{{\tilde z}}
\def\dz{{\sqrt{2}z}}

\def\uij{U_{ij}}
\def\ucij{U^\dagger_{ij}}
\def\uji{U_{ji}}
\def\ucji{U^\dagger_{ji}}
\def\dag{\dagger}

In this letter we would like to show how it is possible to induce a
lattice gauge theory with group $SU(N)$ starting from a lattice
gauge theory for $SU(N)$ coupled to vector in the fundamental irrep.
Differently from (\cite{BaHa,Ar}),
we would like also to consider the relationship among these models,
the dimer problem, branched polymers with loops and "induced QCD" (
\cite{KM}).

We start examining the easiest example, the QED induced, at negative
temperature.
To this purpose let us consider the following action defined on a
lattice $\G$\footnote{
We use $i,j,\dots$ to indicate sites (vertices) of the lattice;
$A,B,\dots$ to indicate rows and colons of group elements $U$;
$\aa,\bb,\dots$ to indicate the generation of the replica .}:
\eq
Z_{U(1)}^{(-)}=
\int~\prod_{links}dU_{i j}\int~\prod_{l\in\G} d\ee_l~d\eb_l~
\exp{\left(-\mu\sum_i\eb_i\ee_i+K\sum_{i,j}A_{i j}U_{i
j}\eb_i\ee_j\right)}
\lbl{u1gho}
\en
where $A_{i j}$ is the adjacency matrix of the lattice $\G$,
$\eb_i=(\ee_i)^*$ are complex grassman variables and
$\uij=e^{i\tt_{i j}}=\ucji$ are elements of $U(1)$.
This action can be regarded as a lattice gauge theory for $U(1)$
with infinite coupling constant $g=\infty$ coupled to a scalar ghost and
because of that it describes a non unitary theory.
The hamiltonian can be rewritten as
\eq
H_{U(1)}^{(-)}=
\sum_{i j}\eb_i(\mu\dd_{i j}-K~A_{i j}\uij)\ee_j
\lbl{h1gho}
\en
and it can be easily verified to be hermitian.
We want to demonstrate that the partition function (\ref{u1gho}) is the
generating function for the the dimer problem, more explicitly:
\eq
Z_{U(1)}^{(-)}=
(-\mu)^N\sum_D~N(D)~(-{K^2\over\mu^2})^D
\en
where $D$ is the number of dimers and $N(D)$ is the number of possible
configurations with $D$ dimers ($N$ is the number of lattice sites).

To prove this assertion we consider the high temperature expansion (HTE)
of the partition function (\ref{u1gho}); to this purpose we give the following
graphical picture to the "interaction terms":

\bigskip
\begin{picture}(200,60)

\put(20,40){\line(1,0){60}}
\put(50,40){\vector(1,0){0}}
\put(20,40){\circle*{1}}
\put(80,40){\circle*{1}}
\put(20,45){$i$}
\put(80,45){$j$}
\put(100,40){$\Longrightarrow~~~K~\eb_i~\uij~\ee_j$}

\put(20,10){\line(1,0){60}}
\put(50,10){\vector(-1,0){0}}
\put(20,10){\circle*{1}}
\put(80,10){\circle*{1}}
\put(20,15){$i$}
\put(80,15){$j$}
\put(100,10){$\Longrightarrow~~~- K~\ee_i~\ucij~\eb_j$}

\end{picture}
\bigskip
\bigskip

%Now in order to have a non vanishing result, we should have on each link
Now if a term of the HTE gives a non vanishing contribution to the
partition function $Z$, it must have
an equal number of $U$ and $U^\dagger$ on each link, but this number
is limited (in
this peculiar case) to 2 since we have grassman variables sitting in
the vertices.
Hence the fundamental unit (dimer) of our expansion is:
\eq
K\eb_i\uij\ee_j \cdot~K\eb_j\ucij\ee_i=
- K^2\eb_i\ee_i~\eb_j\ee_i
\en
which can be given a graphical picture as follows:

\bigskip
\begin{picture}(200,30)

\put(20,11){\line(1,0){60}}
\put(50,11){\vector(-1,0){0}}
\put(20,9){\line(1,0){60}}
\put(50,9){\vector(1,0){0}}
\put(20,10){\circle*{3}}
\put(80,10){\circle*{3}}
\put(20,15){$i$}
\put(80,15){$j$}

\put(100,10){$\equiv$}
\thicklines
\put(120,10){\line(1,0){60}}
\put(150,10){\vector(-1,0){0}}
\put(150,10){\vector(1,0){0}}
\put(120,10){\circle*{1}}
\put(180,10){\circle*{1}}
\put(120,15){$i$}
\put(180,15){$j$}

\put(200,10){$\Longrightarrow~~~-K^2~\eb_i\ee_i~\eb_j\ee_i$}
\end{picture}
\bigskip

Moreover we cannot put two dimers on the lattice sharing a vertex
because $(\eb_i\ee_i)^2=0$, hence we get that the HTE yields just the dimer
recovering problem.
Now we consider the corresponding induced QED. We find easily that
$$
-H_{QED}=N\log(\mu)-
%\sum_{m=1}^\infty{1\oner n}(K\over\mu)^m
\sum_{oriented~\GG
%~with~l(\GG)=n
}{1\over l(\GG)}\left({K\over\mu}\right)^{l(\GG)}~U_\GG
$$
\eq
=N\log(\mu)-
2\sum_{\GG }{1\over l(\GG)}\left({K\over\mu}\right)^{l(\GG)}~Re(U_\GG)
\en
where $\GG$ is a generic closed random walk and $U_\GG$ is the product of the

group element along the oriented path.
Obviously only the non backtracking $\GG$ will give a non trivial result
since $U(1)$ is an abelian group.
Moreover the theory is at negative temperature reflecting the non unitarity
of the initial model, nevertheless this model describes the dimer covering
problem with negative weight that is known (\cite{Sh}) to be equivalent to the
Lee-Yang edge singularity and hence it is critical at $D_{cr}=6$.
We want to stress that the same result could have been reached with the
abelian $\interi_2$ group, i.e. with $U_{i j}=\pm1$ in (\ref{u1gho}).
These results show that the induced $\interi_2$ model and the induced QED
are equivalent each other but they are {\sl not}
equivalent to the usual models at the criticality.
As last observation we want to notice that in the QED case setting $\mu=
0$ we get the close packing dimer problem that it is known to be
equivalent to Ising and hence in this case the theory has $D_{cr}=4$ (
\cite{FiRR}).
%Nevertheless the corresponding induced QED is not in the same
%universality class of the usual QED because, for instance, in $D=3$ it
%is known that the critical exponents of $QED_3$ and of Ising are
%different. There is also another reason of this non equivalence: the same
%result we get for $U(1)$ can be obtained using $\interi$ and hence
%inducing a $\interi$ theory.

In a complete similar way we introduce the induced QED at positive temperature.
It is given by the partition function
\eq
Z_{U(1)}=
\int~\prod_{links}dU_{i j}\int~\prod_{l\in\G}
d\ff_{l}~d\fs_{l}~
e^{
-\mu\sum_i\fs_{i}\ff_{i}+K\sum_{i,j}A_{i j}U_{i j}\fs_{i}\ff_{j}
}
\lbl{u1bos}
\en
where $\ff_i$ are complex scalar defined on the vertices of $\G$.
Its HTE yields
\eq
Z=(w_0)^N\sum_{gas~of~BP}~{\tilde N}(B,\{V_f\})~(K^2)^B~
\prod_{f=1}^{f=q}({w_f\over w_0})^{V_f}
\lbl{zu1bos}
\en
where the sum is extended over a gas of branched polymers with
$B$ bonds, $V_f$ f--branching units and $\tilde N(B,\{V_f\})$ is the
weighted sum of the number of such configurations. Explicitly
$\tilde N(B,\{V_f\})$ is given by
\eq
{\tilde N}(B,\{V_f\})=\sum_{\strut^{\GG~with~B~bonds}_{and~V_f~f~units}}
\prod_{bond~b\in\GG}({1\over n_b!})^2
\en
where $n_b$ is the occupation number of the bond $b$ belonging to a gas
$\GG$ of branched polymers with loops (i.e. $\GG$ can be made of many
disconnected pieces).
Besides the statistical weights of the f--branching units are given by:
\eq
w_f={\pi f!\over\mu^{f+1}}
\en
The formula (\ref{zu1bos}) can be demonstrated
as in the previous case of ghosts: also in this case
a non vanishing contribution to HTE has on each
link an equal number of $U$ and $U^\dagger$, in such a way that
a bond of a branched polymer (a monomer) is composed by one  $U$
and  one $U^\dagger$ bond, exactly as shown by the following figure

\bigskip
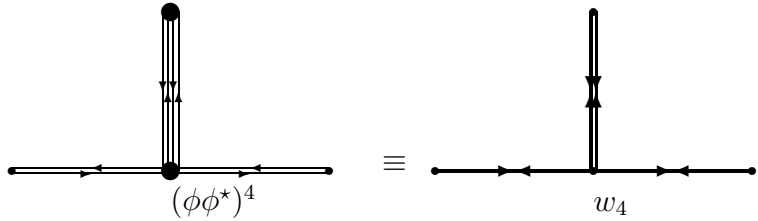
\begin{picture}(300,90)

\put(20,11){\line(1,0){60}}
\put(50,11){\vector(-1,0){0}}
\put(20,9){\line(1,0){60}}
\put(50,9){\vector(1,0){0}}
\put(20,10){\circle*{3}}

\put(80,10){\circle*{7}}
\put(80,-5){$(\ff\fs)^4$}

\put(80,11){\line(1,0){60}}
\put(110,11){\vector(-1,0){0}}
\put(80,9){\line(1,0){60}}
\put(110,9){\vector(1,0){0}}
\put(140,10){\circle*{3}}

\put(77,10){\line(0,1){60}}
\put(79,10){\line(0,1){60}}
\put(81,10){\line(0,1){60}}
\put(83,10){\line(0,1){60}}
\put(77,40){\vector(0,-1){0}}
\put(79,40){\vector(0,1){0}}
\put(81,40){\vector(0,-1){0}}
\put(83,40){\vector(0,1){0}}
\put(80,70){\circle*{7}}

\put(160,10){$\equiv$}
\thicklines
\put(180,10){\line(1,0){60}}
\put(210,10){\vector(-1,0){0}}
\put(210,10){\vector(1,0){0}}
\put(180,10){\circle*{3}}

\put(240,10){\circle*{3}}
\put(240,-5){$w_4$}

\put(240,10){\line(1,0){60}}
\put(270,10){\vector(-1,0){0}}
\put(270,10){\vector(1,0){0}}
\put(300,10){\circle*{3}}

\put(239,10){\line(0,1){60}}
\put(241,10){\line(0,1){60}}
\put(239,40){\vector(0,-1){0}}
\put(239,40){\vector(0,1){0}}
\put(241,40){\vector(0,-1){0}}
\put(241,40){\vector(0,1){0}}
\put(240,70){\circle*{3}}

\end{picture}
\bigskip

In this way each monomer, i.e. polymer link, yields a $\ff\fs$
contribution to each of
its endpoints, so that a f--branching vertex contributes as
\eq
w_f=\int~d\ff d\fs~\exp(-\mu\ff\fs) (\ff\fs)^f={\pi f!\over\mu^{f+1}}
\en
Moreover each link occupied n times yields a factor $({1\over n!})^2$
instead of ${1\over n!}$ because its compositeness.

It is now useful to make a number of comments.
\noindent
Firstly we notice that the this induced QED is also critical at
$D_{cr}=6$. This is easily demonstrated as (see also (\ref{dimd=6})):
$$
Z_{U(1)}\propto
\int\prod_{links}dU_{i j}\det(\mu\dd_{i j}-K~A_{i j}\uij)^{-1}
=\lim_{n\rightarrow~-1}\int\prod_{links}dU_{i j}
\left(\det(\mu\dd_{i j}-K~A_{i j}\uij)\right)^{n}
$$
\eqa
=\lim_{n\rightarrow~-1}\int~\prod_{links}dU_{i j}\int~\prod_{l\in\G}
d\ee_{l \aa}~d\eb_{l \aa}~
e^{\sum_{\aa=1}^n\left(
-\mu\sum_i\eb_{i \aa}\ee_{i \aa}
+K\sum_{i,j}A_{i j}U_{i j}\eb_{i \aa}\ee_{j \aa}
\right)}
\nonumber\\
{}~~
\ena
and relying on the fact that
the critical dimension is independent of the number of components
of the order parameter.

\noindent
Secondly the difference between branched polymers with loops and random
walks is {\sl both} in the silhouette of the terms of the HTE
( due to the appearance of odd branching vertices) and
in the weights used for $\tilde N$ ($ 1\over n_b!$ for RW,
$({1\over n_b!})^2$ for BP).
%These two remarks show that the actual understanding of branched
%polymers is far from being complete and necessitate of further
%investigations.

Now we would like also to consider the more general case of the
$SU(N)$ or $U(N)$ induced by the the following partition function:
\eq
Z_{SU(N)}
=\int~\prod_{links}dU_{i j}\int~\prod_{l\in\G}
d\ff_{l A}~d\fs_{l A}~
e^{-\mu N\,
\sum_i\fs_{i A}\ff_{i A}+N K\,\sum_{i,j}A_{i j}U_{i j,A B}
\fs_{i A}\ff_{j B}}
\lbl{unbos}
\en
where $\ff_i\equiv(\ff_{i A})=(\ff_{i 1}\dots\ff_{i N})$ are complex
scalars in the fundamental irrep of $SU(N)$ defined on every vertex
$i\in\G$.
The HTE of these models  yield graphs with the same silhouettes of the
$U(1)$ case but with different weights.
To get these weights we integrate over the gauge variables firstly, to
this purpose we use the fact that
$$
f_{SU(N)}(M,M^\dag)=e^{N\,g_{SU(N)}(M,M^\dag)}=
  \int_{SU(N)}~dU_{i j}~exp(N\,tr(M^\dag U+U^\dag M))
$$
is a function of $tr(M M^\dag)^p$ and of $det M$ (in the case of
$SU(N)$) and that in our case $M_{i j}=\ff_i\otimes\fd_j$ to deduce that
$tr(M M^\dag)^p_{i j}=(\fd\ff)^p_i~(\fd\ff)^p_j$ and $det M=0$,
so that\footnote{It is easy to show that using the normalization $vol(U(
N))=vol(SU(N))$ we have $f_{U(N)}(\ff_i,\ff_j)=f_{SU(N)}(\ff_i,\ff_j)$.
Moreover this implies that using $N>1$ scalars we cannot induce $U(N)$
but only $SU(N)$.}
\eq
f_{SU(N)}(\ff_i\otimes\fd_j,\ff_j\otimes\fd_i)
=e^{N\,g_{SU(N)}(\fd_i\ff_i\times\fd_j\ff_j)}=
\sum_p c_p(N)~(\fd\ff)^p_i~(\fd\ff)^p_j
\lbl{fun}
\en
Hence we can rewrite (\ref{unbos}) as
\eq
Z_{SU(N)}=
\int~\prod_{l\in\G}
d\ff_{l A}~d\fs_{l A}~
e^{-N\mu
\sum_i\fd_{i }\ff_{i }
}~\prod_{links~ i j}
\left(\sum_p c_p(N)~K^{2p}(\fd\ff)^p_i~(\fd\ff)^p_j\right)
\lbl{unbos1}
\en
Similarly to the $U(1)$ case we associate to each polymer link a factor
$\fd\ff$ for each of its endpoints, so that now the weight for a
f-branching vertex becomes
\eq
w_f=\int~\prod_A d\ff_A d\fs_A~\exp(-\mu N\fd\ff) (\fd\ff)^f
={\pi^N (N+f)!\over N!(\mu N)^{N+f+1}}
\en
meanwhile the occupation weights can be deduced from (\ref{fun}) and
yield for ${\tilde N}(B,\{V_f\})$ the expansion
\eq
{\tilde N}(B,\{V_f\})=\sum_{\strut^{\GG~with~B~bonds}_{and~V_f~f~units}}
\prod_{bond~b\in\GG}c_{n_b}(N)
\en
where in particular we have $c_n(1)=({1\over n!})^2$.

It is important to notice that also these models are critical at $D_{cr}
=6$ because
they are in the same universality class of branched polymers with loops
and both odd and even vertices.
In order to show this point also from a more mathematical point of view,
we change coordinates in (\ref{unbos1}) and we introduce
$\exp{\vf_i}=\fd_i\ff_i$ ($\vf\in]-\infty,+\infty[$) along with
the radial coordinates, so that (\ref{unbos1}) becomes
\eq
Z_{SU(N)}= const\,
\int~\prod_{l\in\G}
d\vf_l
\exp\left(
N\sum_i(\vf_i-\mu e^{\vf_i})
+N\sum_{<i j>}g_{SU(N)}(e^{\vf_i+\vf_j})
\right)
\lbl{dimd=6}
\en
If now we expand the action, we shift $\vf$ as the shifted $\vf$ has
null expectation value and we throw away the irrelevant term we recover
a $\vf^3$ action that is critical at $D_{cr}=6$. Obviously this is not a
rigorous proof since we should check that the irrelevant operators are
really irrelevant at the fixed point (\cite{FP}), nevertheless up to now
this approach did not fail to get the upper critical dimension.

Extending the known result valid for ordinary lattice gauge theories
for which both $SU(N)$ and $SU(N)/\interi_N$ have the same upper
critical dimension, here we would deduce that both the vector induced
lattice gauge theories and the adjoint induced theories ("induced QCD")
at finite $N$ share the same upper critical dimension $D_{cr}=6$,
nevertheless this seems not to be true in this case since the induced
QCD seems to be critical at $D_{cr}=4$ (\cite{CDP}).

Now we would solve exactly this model in the large $N$ limit. To this
purpose we compute exactly (\ref{fun}) in the large $N$ limit.
Using the observation that $f_{SU(N)}$ depends only on
$x=\fd_i\ff_i\cdot\fd_j\ff_j$ we get easily the equation
\eq
{x\over N}{d^2 g\over d x^2}
+({d g\over d x})^2
+{d g\over d x}=1
\lbl{eq-fun}
\en
that can be solved easily (\cite{BG}) as:
\eq
g(x)=\sqrt{1+4 x}-1 -\log(1+\sqrt{1+4 x})+log2
\lbl{sol-fun}
\en
Plugging this solution into (\ref{unbos1}) and changing coordinates from
$\ff_{i A}\,\, ,\fs_{i A}$ to $R_i=\fd_i\ff_i$ and angular coordinates,
we get
\eq
Z_{SU(N)}=
const\,\int_0^\infty\,\prod_{i\in\G}\,dR_i\,
e^{N\,\left[
\sum_i \left(1-{1\over N}\right)\,\log(R_i)-\mu\,R_i
+\sum_{<i j>}\sqrt{1+4 K^2R_iR_j}-\log(1+\sqrt{1+4 K^2R_iR_j})
\right]}
\en
Now we can use the usual saddle point method. If we look for a
translational invariant solution  we get
\eqa
K\,R&=&{{ -{\mu\over K}(D-1)+\sqrt{ ({\mu\over K})^2-4(2D-1) } }\over
        { ({\mu\over K})^2-4D^2}  }
{}~~~~~~\mbox{for ${\mu\over K}>2D$ and $D\ge0$ }
\nonumber\\
K\,R&=&{2D-1\over 4D(D-1)}
{}~~~~~~\mbox{for $D>1$}
%\nonumber\\
%K\,R&=&?
%~~~~~~\mbox{for $D<{\mu\over J}<2D$ }
\ena
while in the range $D<{\mu\over J}<2D$ the saddle point is unstable.
The explicit form of the free energy density is then ($\cal V$ is the
volume of the lattice)
\eq
F_{SU(\infty)}={\log Z_{SU(N)}\over N {\cal V}}=
\log(KR)-{\mu\over J}R
+D\sqrt{1+4 K^2R^2}-D\log(1+\sqrt{1+4 K^2R^2})
-\log(K)
\en
that does not show any transition. It can be shown that
$F_{SU(\infty)}^{\mbox{fer}}=-F_{SU(\infty)}^{\mbox{bos}}$.

In conclusion we show that the $N=\infty$ theories (both bosonic and
fermionic) are "trivial" since they have no critical dimension and
can be solved exactly by the saddle point; on the contrary the theories
at $N\ne\infty$ are critical so the role of coupling constant is played
by $1\over N$ but the $N=\infty$ theory is not able to catch all the
features of the $N\ne\infty$ theories, exactly as the gaussian
approximation has not all the characteristics of $\ff^4$ theory.
Perhaps the same conclusion can be reached for the $SU(N)$ induced by up
to $N-1$ scalars since in this range we are in the strong coupling regime
of Brezin-Gross formula (\cite{BG}), nevertheless a different regime
should show up starting at $N$ scalars (or fermions) since it appears
both the strong and weak coupling regime (\cite{Ar}).

{\bf Acknowledgement}
We thank for useful discussion D. Boulatov, M. Caselle, A. D'Adda,
P. Di Vecchia, F. Gliozzi and M. Rasetti.

\end{document}